\documentclass[aps,prd,showpacs,floatfix,nofootinbib,twocolumn,10pt]{revtex4-1}
\usepackage{color,amsmath,amssymb,graphicx,latexsym,subfigure}
\usepackage{soul,color}
\usepackage[dvipsnames]{xcolor}

\usepackage{tabularx} 
\usepackage{amsmath}  
\usepackage{graphicx} 
\usepackage[final]{hyperref} 
\hypersetup{
	colorlinks=true,       
	linkcolor=blue,        
	citecolor=blue,        
	filecolor=magenta,     
	urlcolor=blue         
}

\newcommand{\beq}{\begin{equation}}
\newcommand{\eeq}{\end{equation}}
\newcommand{\bea}{\begin{eqnarray}}
\newcommand{\eea}{\end{eqnarray}}

\usepackage[arrows]{ezedits}
\defineEdit{AH}{\color[rgb]{1.0,0.0,0.0}}{\color[rgb]{1.0,0.0,0.0}}
\defineEdit{RQ}{\color[rgb]{0.63,0.12,0.94}}{\color[rgb]{0.63,0.12,0.94}}

\begin{document}

\title{Gravitational focusing effects on streaming dark matter as
a new detection concept }


\author{Abaz Kryemadhi$^{a}$, Marios Maroudas$^{b}$, Andreas Mastronikolis$^{c}$, and Konstantin Zioutas$^{b}$
}
\affiliation{
$^{a}$ Dept. Computing,Math \& Physics, Messiah University, Mechanicsburg PA 17055, USA
\\
$^{b}$ Physics Department, University of Patras, GR 26504,Patras-Rio, Greece
\\
$^{c}$ Dept. Physics \& Astronomy, University of Manchester, Manchester M13 9PL, UK 
}

\begin{abstract}

Cosmological simulations for cold dark matter (DM) indicate that a large number of streams might exist in our Galaxy.
The present work incorporates gravitational focusing (GF) effects on streaming DM constituents by the Sun and the Earth preceding their encounter with Earth bound detectors.    
For streaming DM, the GF gives rise to spatiotemporal flux enhancements of orders of magnitude above the nominal DM density.
Remarkably, due to Earth’s rotation  the derived flux enhancements appear as transient signals lasting about 10 seconds repeating daily for days or weeks. 
This work presents a novel opportunity for DM signal detection and identification, and 
the present simulation can be applied to any kind of invisible matter entering the solar system.

\end{abstract}

\maketitle
 
\section{Introduction}



Dark matter constitutes about 80\% of the matter in the universe and plays an important role in structure formation. In the Standard Halo Model (SHM), DM is described as a cold collision-less fluid with a smooth Maxwellian velocity distribution ($v \approx 0.001\,\text{c}$, where c is the speed of light), which is an approximation since cold DM simulations indicate non smooth features~\cite{Vogelsberger:2020,Stucker:2020}.  

Much progress has been made in studying the small scale structure of DM in the Galaxy, particularly the velocity distribution in the halo despite the N-body computational challenges related to finer resolution requirement~\cite{Springel:2021,Vogelsberger:2020}. The study of small-scale structure of DM spanning the size of solar system is important, as it can shed light on the local DM density distribution and, eventually, on the flux of particles at Earth bound experiments~\cite{Vogelsberger:2009,Vogelsberger:2011}.

Vogelsberger et. al.~\cite{Vogelsberger:2009,Vogelsberger:2011} used a geodesic deviation equation and the N-body equations of motion to study the evolution of phase space of the DM halos. They concluded that many caustics and fine grained streams could be present in our solar system. Fine grained streams are a consequence of cold and collisionless DM, which is restricted to a 3D hypersurface in a 6D phase space. The thickness of this surface reflects the primordial velocity dispersion~\cite{Vogelsberger:2008, Vogelsberger:2011}.  
We draw attention to the fact that the fine grained streams studied in this work are different from tidal streams, which are caused by halo disruptions~\cite{Gaia:2017}. The simulations carried out by Vogelsberger et al.~\cite{Vogelsberger:2011} for fine grained streams indicate that up to $\sim10^{12}$ such streams could be present in the vicinity of the solar system, and about $10^{6}$ streams contain half the average local DM density. Consequently, the DM content at a point in the halo is described as the superposition of many fine-grained streams, each with a small velocity dispersion. 
The velocity dispersion is about $\sim 10^{-17}$c for fine grained axions, and about $\sim 10^{-10}$c for WIMPs~\cite{Vogelsberger:2009,Vogelsberger:2008, Sikivie:1995,SikivieVelo:1999,natarajan:2008}.

The small dispersion is attractive for axion haloscopes due to the increased coherence length and time with smaller velocity dispersion. This appears as a quality factor benefit, attributed to the fact that the axion to photon conversions in a cavity will grow coherently over a large coherence time. (See Refs.~\cite{slee:2020,admx:2018,CAST:2017})

In this work, we explore these small velocity dispersion streams by studying the density enhancements caused by the GF effect of the intervening solar system bodies.

Gravitational focusing effect of solar system bodies on DM has been subject of many studies.  In the subsequent discussion, we summarize some of these findings.   
Griest~\cite{Griest1988} studied the GF effect by the Sun on DM particles in the vicinity of the Earth, and identified an annual modulation of about 1\%. The modulation was found to be comparable to the annual modulation caused by the relative velocities of the Sun and the Earth. Bozorgnia and Schwetz~\cite{Bozorgnia_2014} examined the impact of GF on the annual modulation signal in direct detection experiments. They found that the GF-induced modulation signal is relatively small. When the GF effect was taken into account, the parameter space was reduced compared to the scenario where the GF effect was not considered.  Similarly, Lee et al.~\cite{lee:2014} investigated the GF effect by the Sun on the annual modulation signal. They determined that the GF effect is limited to a maximum of 3\%. However, they highlighted a significant overall shift in the phase of annual modulation, which is particularly relevant for DM particles with low scattering speeds. Kouvaris and Nielsen~\cite{Kouvaris2015} conducted a study on the GF effect of the Earth on annual modulation. Their findings indicated that GF has the potential to generate a more prominent diurnal modulation compared to the modulation caused by the Earth's rotation around its own axis. 
Sikivie and Wick~\cite{SikivieWick2002} studied the GF effects by the Sun on a cold dispersion-less flow of dark matter. They concluded that regions with quasi-infinite density (caustics), appear downstream of the Sun. Furthermore, Alenazi and Condolo~\cite{alenazi:2006} examined the distribution of unbound collision-less flow of DM particles in the Solar System using numerical and analytical methods. They specifically focused on the velocity distribution at Earth's location and demonstrated that particularly for a flow, rings are formed in the arrival distribution of DM when Earth is positioned behind the Sun as seen from the flow.
Nobile, Gelmini, and Witte~\cite{Nobile_2015} analyzed the GF effect by the Sun on a halo characterized by the standard SHM. Their study encompassed the consideration of a tidal stream (Sagittarius) and also incorporated a dark disk component. The findings indicated that the GF effect tends to diminish certain characteristics in the amplitudes and phases of the annual and biannual harmonics.

Hoffmann, Jacoby, and Zioutas~\cite{Hoffmann:2003}, studied the GF effects by the Sun on streaming non relativistic DM particles and concluded that the flux can be temporarily amplified at the site of the Earth with effective DM flux amplification factors by as much as $10^3$ to $10^4$. Building upon this work, Patla et al.~\cite{Patla:2014} expanded the analysis to include substantial GF effects by Jupiter on particles with speeds ranging from $0.01$c to $0.001$c.
Prezeau~\cite{Prezeau:2015} employed the geodesic equation to examine the GF effects of Earth and Jupiter on streaming DM particles. The study specifically focused on streams of DM with a nominal velocity of 220 km/s and investigated the feasibility of detecting regions of high density. However, the detection of these high density regions presents a challenge due to the large distance between them and the Earth. Prezeau introduced the term "DM hairs" to describe these spatial areas showing notable enhancements in the DM density~\cite{Prezeau:2015}.  Sofue~\cite{sofue:2020} examined the phenomenon of self-focusing caused by the gravity of the Earth on low-speed DM particles. The findings revealed that dispersion-less streams of low-speed DM could produce significant density enhancements regions in close proximity to the Earth.

As a summary of the aforementioned studies, we highlight that the GF effects, assuming a smooth SHM, lead to a flux variation of a few percent and occurring at specific times of the year. However, in the case of streaming DM scenarios as we address in this simulation, the enhancements can reach several orders of magnitude, making them potentially significant for DM detection (as discussed in Refs.~\cite{Hoffmann:2003, Patla:2014, Prezeau:2015, sofue:2020}).


 The potential of GF to cause significant density enhancements for low-velocity streams, which could be observed through both space and ground-based experiments, is a compelling motivation for our simulation study.
 The present work extends to include the combined GF influence of both, the Sun and the Earth.
 Moreover, it takes into account the Earth's inclined ecliptic plane with respect to galactic plane which imposes kinematical constrains for streaming DM particles that reach the Earth. Furthermore, the dispersion effects on density enhancements are also incorporated. Quantum effects such as those for very light bosonic DM are not considered~\cite{Kim_Lenoci_2022}. We pivoted the study on fine grained streams of axions and WIMPs, however the simulation could be used broadly to evolve the trajectories of any type of DM particle under the gravitational influence of the Sun and the Earth. In our analysis, we will refer to regions with enhanced density due to gravitational focusing influence as DE regions (Density Enhancement regions). 

 This paper is organized as follows: Section~\ref{sec:simulation} describes the simulation approach used, followed by the results for the Sun Earth system in section~\ref{sec:focusing}. We then discuss the prospect for detection of these DE regions of different densities in~\ref{sec:discussion}, and the summary with conclusions is outlined in section~\ref{sec:summary}. 
 
\section{Simulation}
\label{sec:simulation}

The performed simulation begins with an initial phase-space distribution of DM, which encompasses both positional and velocity distributions in the galactic rest frame. It incorporates various dispersion profiles, ranging from dispersionless cases, those for axions, WIMPs, or the overall SHM, as input parameters. The simulation then evolves the phase-space (velocity and spatial distribution) of DM particles in the gravitational field of the Sun and the Earth and the resulting distributions near the Earth are then obtained.

\vspace{3mm}
{\bf {Initial phase-space of the simulation}}
\vspace{1mm}

It is important to acknowledge that current cosmological simulations do not have solar or sub-solar system resolution scales~\cite{Angulo:2022, Vogelsberger:2020}. Consequently, in the absence of input from such simulations, our work makes the reasonable assumption that the initial spatial distribution of DM particles in each stream is uniform and that the streams span a size at least as large as the solar system.

In the smooth SHM, the velocity distribution in the galactic rest frame is presumed to follow a Maxwellian distribution with a standard deviation of $\sigma_\text {v}\approx {(230/\sqrt{2})\,\text{km/s}} \approx 160\,\text {km/s}$. However, for our purposes, the velocity distribution of streams is required instead.

\begin{figure}[htbp]
  \centering
  \includegraphics[width=3.2in]{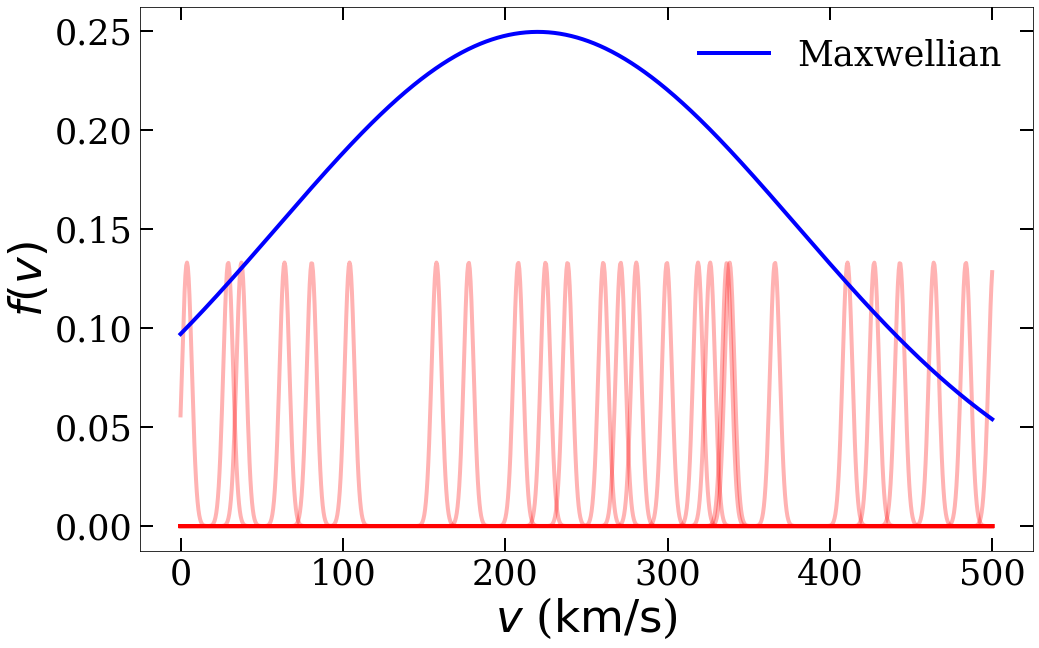}
  \caption{The process of selecting velocities of streams in the simulation. The selection of velocities is optimized to have a higher concentration of streams near the peak of the Maxwellian distribution, with fewer streams selected towards the tails of the distribution. The plot depicts a limited number of streams with their widths being exaggerated to enhance visibility and clarity. }
  \label{fig:velocity_show}
\end{figure}

References~\cite{Sikivie:1995,Vogelsberger:2008} suggest that a fraction of cold DM thermalizes by undergoing numerous orbits in the Milky Way, resulting in broadening of the velocity distribution. However, the exact fraction of cold DM that has undergone sufficient orbits remains unknown. Cold DM with fewer orbits would exhibit sharper peaks and a smaller velocity dispersion~\cite{Sikivie:1995,Vogelsberger:2008}. Since the precise velocity distribution of each stream is uncertain, we suppose that $10^{12}$ discrete streams with different velocities superimpose in a manner that reproduces the overall Maxwellian velocity distribution, with a mean local DM density of $\rho_{\text 0}=0.45\,\text{GeV} \text{cm}^{-3}$~\cite{Vogelsberger:2011}. Given this assumption, we expect that combining the velocity distributions from multiple streams will result in an overall distribution that closely approximates a Maxwellian distribution. To visualize this process, we provide Fig.~\ref{fig:velocity_show}, which demonstrates how stream velocities are chosen in the simulation and their respective abundances. The selection of velocities is designed to favor a larger number of streams near the peak of the Maxwellian distribution, while fewer streams are chosen towards the tails of the distribution. It is important to acknowledge that even though the resulting overall distribution may appear continuous, it is reconstructed from discrete streams.

Unsurprisingly, the total velocity conforms to the Maxwellian distribution as per our assumption. This collective velocity arises from the combined effects of multiple stream velocity distributions. It can be mathematically expressed as follows:
\begin{equation}
\label{eq:fvtot}
f_{\text {tot}}(\bf v_{\text {g}}) =\frac{1}{N} \sum_{k} N_{\text {sk}}\delta^{3} (\bf v_{\text {g}} -\bf v_{\text {sk}}),    
\end{equation}
where $\delta^{3} (\bf v_{\text {g}} -\bf v_{\text{sk}})$ is the Dirac-delta function, $\bf v_{\text {sk}}$ is the velocity of k-th stream, ${\textbf v_{\text {g}}}$ refers to the velocity of DM particles in the galactic rest frame, $\bf{N}$ is the total number of particles, and ${\bf N}_{\text {sk}}$ is the number of particles in that particular stream. The Maxwellian velocity distribution function $f_{\text {tot}}(\bf v_{\text{g}})$ is implemented as a multivariate Gaussian with parameters in the galactic rest frame given by Table~\ref{tab:simulation_table}. 
To implement the Dirac-delta functions, we utilize narrow Gaussians with a width that represents the dispersion of the respective stream. It is important to note that for dispersionless streams, particles within the same stream possess identical velocities.

\begin{table}
	\centering
	\caption{Input simulation parameters in galactic frame. A multivariate Gaussian distribution is used for the velocity profile with a mean value of 0 and $\sigma$ being the same for radial, tangential, and z-direction.}
	\label{tab:simulation_table}
	\begin{tabular}{lc} 
		\hline
		Parameter & SHM \\
		\hline
		$\sigma_r$ & 160~km/s\\
		$\sigma_{\phi}$ & 160~km/s\\
		$\sigma_{z}$ & 160 km/s\\
		$v_{\phi}$ & 0~km/s\\
		$v_{r}$  & 0~ km/s\\
		$v_{z}$  & 0~km/s \\
		\hline
	\end{tabular}
\end{table}

\vspace{3mm}
{\bf {Gravitational effect by the Sun}}
\vspace{1mm}

\begin{figure}[h!]
  \includegraphics[width=3.2in]{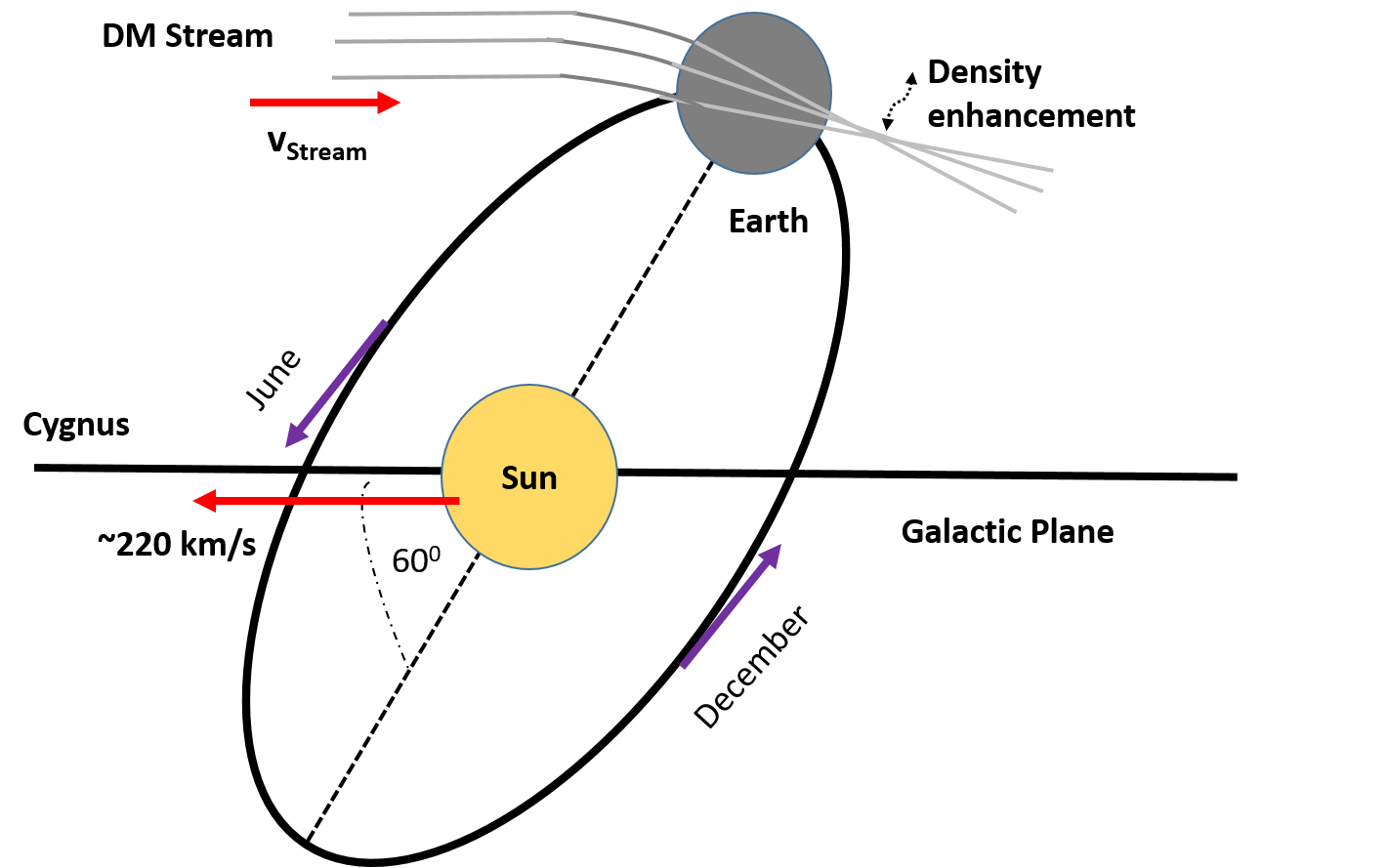}
\caption{Schematic illustration of DM particles deflected gravitationally by the Sun and then self-focused by the Earth. }
\label{fig:schematics}
\end{figure}

After initializing both spatial and velocity distributions of the streams, the simulation proceeds to calculate the trajectories of DM particles under the influence of the gravitational fields exerted by the Sun first and then the Earth. In Fig.~\ref{fig:schematics}, the schematic shows the trajectories of DM particles from a stream as they fall into the gravitational well of the Sun, followed by their subsequent focusing on the opposite side of the Earth. This focusing effect is a result of the gravitational force exerted by the Earth while the DM particles propagate inside the Earth towards its opposite side.

Given ${\textbf v_{\text {g}}}$ as the velocity of DM particles in the galactic rest frame, the velocity of DM particles relative to the Sun (denoted as ${\textbf v_{\infty}}$) when situated at a considerable distance from the Sun, can be expressed as:
\beq
\label{eq:vinf}
{\textbf v_{\infty}} ={\textbf v_{\text {g}}-\textbf v_{\odot}},  
\eeq
where $\textbf v_{{\odot}}$ is the Sun's velocity in galactic rest frame and it is determined by the sum of its peculiar velocity and the velocity of the Local Standard of Rest (LSR). The peculiar velocity, represented by $\textbf{v}_{\text{pec}}$ relative to the LSR, is specified as $(11, 12, 7)$ km/s, while the velocity of the LSR, denoted as $\textbf{v}_{\text{LSR}}$, is given as $(0, 230, 0)$ km/s relative to galactic rest frame according to Refs.~\cite{McMillan:2017} and \cite{chare:2014}. This relationship can be expressed as: 
\beq
\label{eq:vsun}
{\textbf v_{\odot}=\textbf v_{\text {pec}} + \textbf v_{\text {LSR}}},
\eeq



DM particle velocities at a considerable distance from the Sun are represented by ${\textbf v_{\infty}}$. As they approach the Earth, the Sun's gravitational effect modifies their velocity, which is denoted as ${\textbf v}$. By applying the principles of conservation of energy and the Laplace-Runge-Lenz vector, the following relationship can be established (for more details, refer to Refs. ~\cite{alenazi:2006, chare:2020}).

\beq
\label{eq:velocity}
{\textbf v_{\infty}} =\frac{\text v_\infty^2{\textbf v}+{\text v_\infty}(GM/r){\hat{\bf {r}}}-{\text v_\infty}
{\textbf v}\left({\textbf v}\cdot{\hat{\textbf r}} \right)}
{ {\text v_\infty^2}+\left(GM/r\right)-{\text v_\infty}\left(\textbf{v}\cdot{\hat{\textbf r}}\right) },  
\eeq
where 
\beq
v_{\infty} =\sqrt{v^2-\frac{2 G M}{r}}
\eeq
relates the speeds. We used python's {\it scipy.optimize.fsolve} function to derive ${\textbf v}$ from Eq.~\ref{eq:velocity} given ${\textbf v_{\infty}} $ from Eq.~\ref{eq:vinf}.

At this stage, ${\textbf v}$ represents the velocity of DM particles relative to the Sun. To obtain their velocity relative to the Earth (${\textbf v_{\text {lab}}}$), we must consider Earth's rotational velocity (${\textbf V_{\oplus}}$) around the Sun. This can be expressed as:
\beq
\label{eq:vlab}
{\textbf v_{\text {lab}}=\textbf v-\textbf V_{\oplus}}.
\eeq
To position the Earth with respect to the Sun in the Galaxy, we used a code developed by C.~O'Hare~\cite{chare:2014}.  

By solely considering the gravitational influence of the Sun, we can now examine the spatial and velocity distribution of DM particles.

\begin{figure}[h!]
\includegraphics[width=1\linewidth]{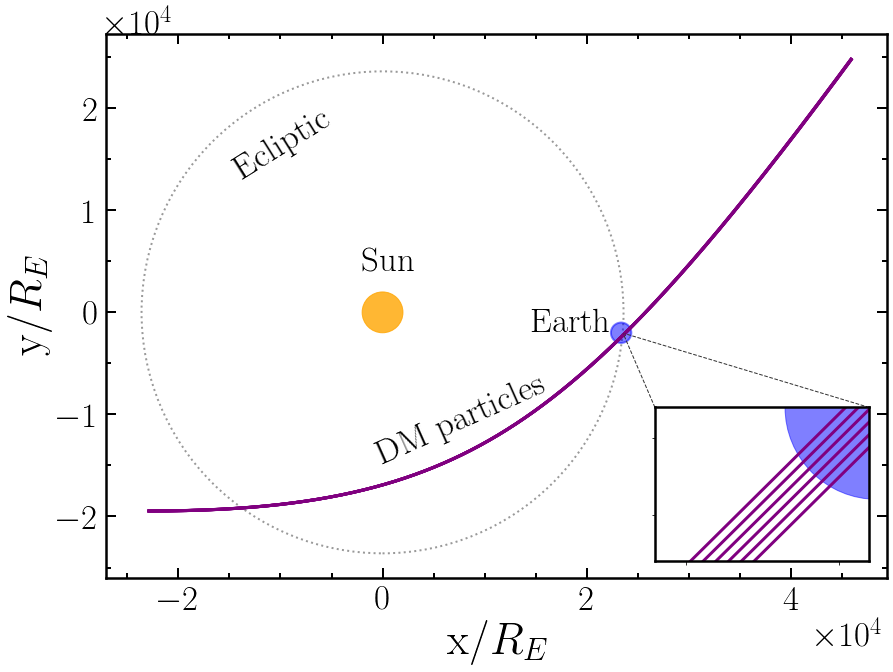}
\caption{The illustration depicts the impact of the Sun's gravity on a DM stream. As the particles fall into the gravitational well of the Sun, their direction is modified. However, upon closer examination in the zoomed-in plot at the bottom right corner, it is evident that the spatial uniformity of the Earth-bound stream particles remains mostly unaffected by the Sun's influence. The Earth and the Sun sizes are not drawn to scale. }
\label{fig:sun_only_gf}
\end{figure}

Figure~\ref{fig:sun_only_gf} demonstrates the Sun's gravitational impact on DM particles from a stream. As these particles approach the Earth, their overall direction undergoes a change as described by Equation~\ref{eq:velocity}. However, upon zooming in near the Earth, it becomes evident that these Earth-bound DM particles (particles within the size of the Earth's aperture) maintain spatial uniformity, indicating a negligible tidal effect from the Sun. Consequently, the Sun modifies the direction of the streams while maintaining the low velocity dispersion of Earth-bound DM particles.  Importantly, these modifications by the Sun do not lead to significant density enhancements for these DM particles.


\begin{figure}[h!]
\includegraphics[width=1\linewidth]{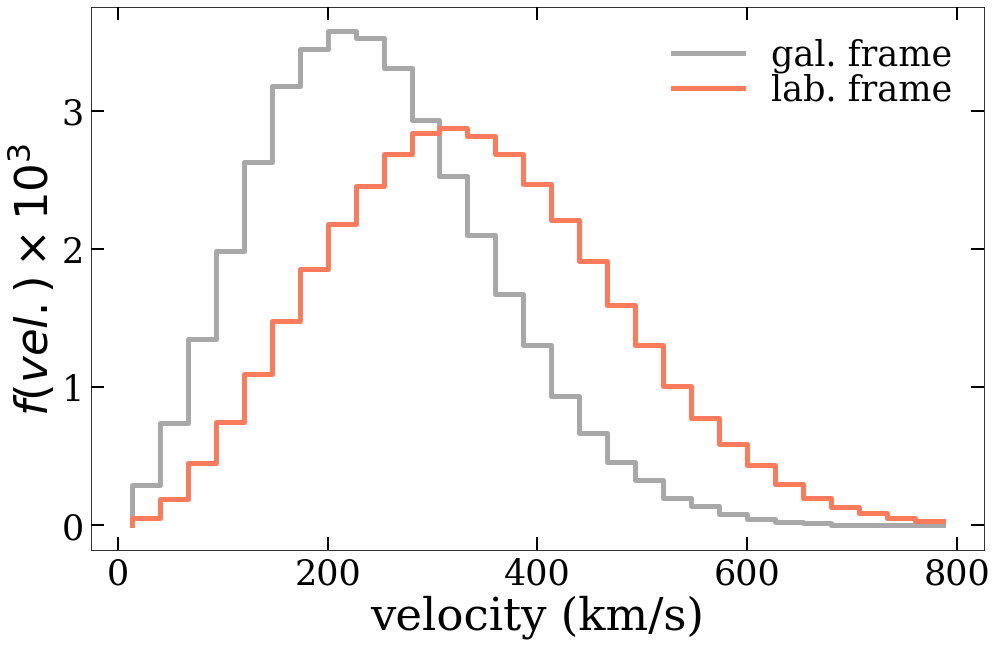}
\caption{The velocity distribution of DM particles for a large number of streams in the galactic and laboratory frames. The laboratory frame exhibits a higher mean velocity compared to the galactic frame due to the kinematic boost and the gravitational effect by the Sun. The coarse binning obscures the discrete streaming nature of the distribution, and the Sun's gravitational effect does not create distinct narrow peaks at a specific velocities. This is due to the Sun mainly altering the direction of the streams rather than generating density enhancements.  }
\label{fig:velocities}
\end{figure}

Figure~\ref{fig:velocities} displays the velocity distribution of DM particles from multiple streams in both the galactic and laboratory frames. As anticipated, the mean velocity is higher in the laboratory frame compared to the galactic frame. The velocity distribution in the laboratory frame incorporates the kinematic boost and the gravitational effect exerted by the Sun, as described by Eqs.~\ref{eq:vinf},~\ref{eq:velocity}, and~\ref{eq:vlab}. The coarse binning of the plot obscures the discrete streaming nature of the distribution. Furthermore, the inclusion of the Sun's gravitational effect does not result in a distinct narrow peak at a particular velocity. This is because the Sun mainly alters the direction of the Earth bound streams and does not produce significant density enhancements as it is shown for the general configuration in Figs.~\ref{fig:schematics} and ~\ref{fig:sun_only_gf}.

\vspace{3mm}
{\bf {Gravitational effect by the Earth}}
\vspace{1mm}

After the Sun's gravitational influence, the streams approach near the Earth and experience its gravitational effect.
The DM trajectories are reconstructed using Newton's Second Law with the python's {\it scipy.odeint} package. The equations are:
\beq
 {\text{\bf v}_{\text{i}}}=\frac{d\text{\bf r}_\text{i}}{\text{dt}}
\eeq
\beq
\label{eq:Earth}
\frac{\text {d}{\bf v_i}}  {\text {dt}} =\begin{cases}
-\text {G}\frac{\text {M}_{ {E}}}  {\text {r}_i^3} {\bf r_i} & ~~~~\text {r}_i\geq  {R}_{{E}} \\[0.5em]

-\text{G}\frac{4 \pi \eta_{{E}} (\text{r}_i)}  {3} {\bf r_i} &  ~~~~ \text{r}_i< {R}_{{E}},  \\
\end{cases}
\eeq


where $\text{M}_{E}$, $R_E$ are the mass and radius of the Earth, ${\bf r_i}$ and ${\bf v_i}$ are the position and velocity of the $i^{th}$ DM particle, and $\eta_{E} (\text{r}_i)$ is the local Earth density. The center of the Earth is located at the origin of the reference frame. For the Earth's density, an approximate PREM model~\cite{prem:1981} is used, with densities $12000\,\text {kg}/\text{m}^3$, $7000\,\text{kg}/\text{m}^3$, and $5000\,\text {kg}/\text {m}^3$, for inner distances from 0-3400 km, 3400-5700 km, and from 5700-6340 km, respectively. Throughout this work a possible self-interaction of DM particles is ignored. As depicted in Fig.~\ref{fig:focusing}, the DM particles from a stream converge towards a focal region after passing through the Earth where a high-density region occurs.   


\vspace{3mm}
{\bf {Combined gravitational effect (Sun and Earth)}}
\vspace{1mm}

Our current approach involves the utilization of Eq.~\ref{eq:velocity} to determine the velocity distribution of DM particles from streams, taking into account the gravitational influence of the Sun which as per Fig.~\ref{fig:sun_only_gf} mainly alters the direction of Earth bound DM particles in the stream without causing density enhancement. Furthermore, Eq.~\ref{eq:vlab} is applied to convert the velocity into the laboratory frame of reference. Afterwards, DM particles from the same stream, located at an approximate distance of $\sim30\,\text{R}_{\text{E}}$ from the Earth, are introduced into the Earth's gravitational field using Eq.~\ref{eq:Earth}. This distance is considered to be sufficiently far from the Earth for the purposes of our study. The density enhancement, which we study in the subsequent sections, arises solely from Earth's gravitational influence.

\section{DM Flux Enhancements}
\label{sec:focusing}

\vspace{3mm}
{\textbf {Amplification}}
\vspace{1mm}

\begin{figure}[h!]
 \includegraphics[width=1\linewidth]{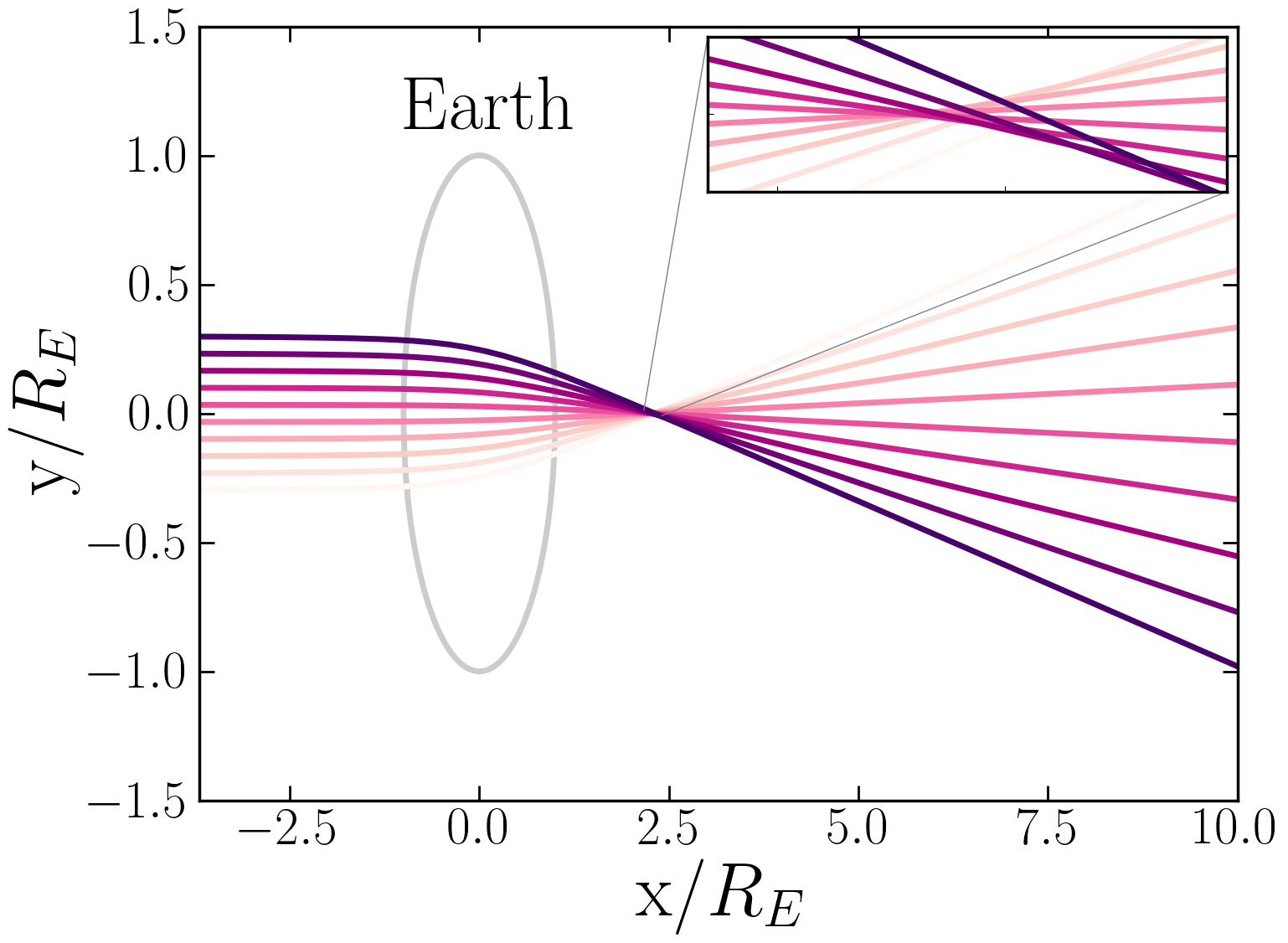}
  \includegraphics[width=1\linewidth]{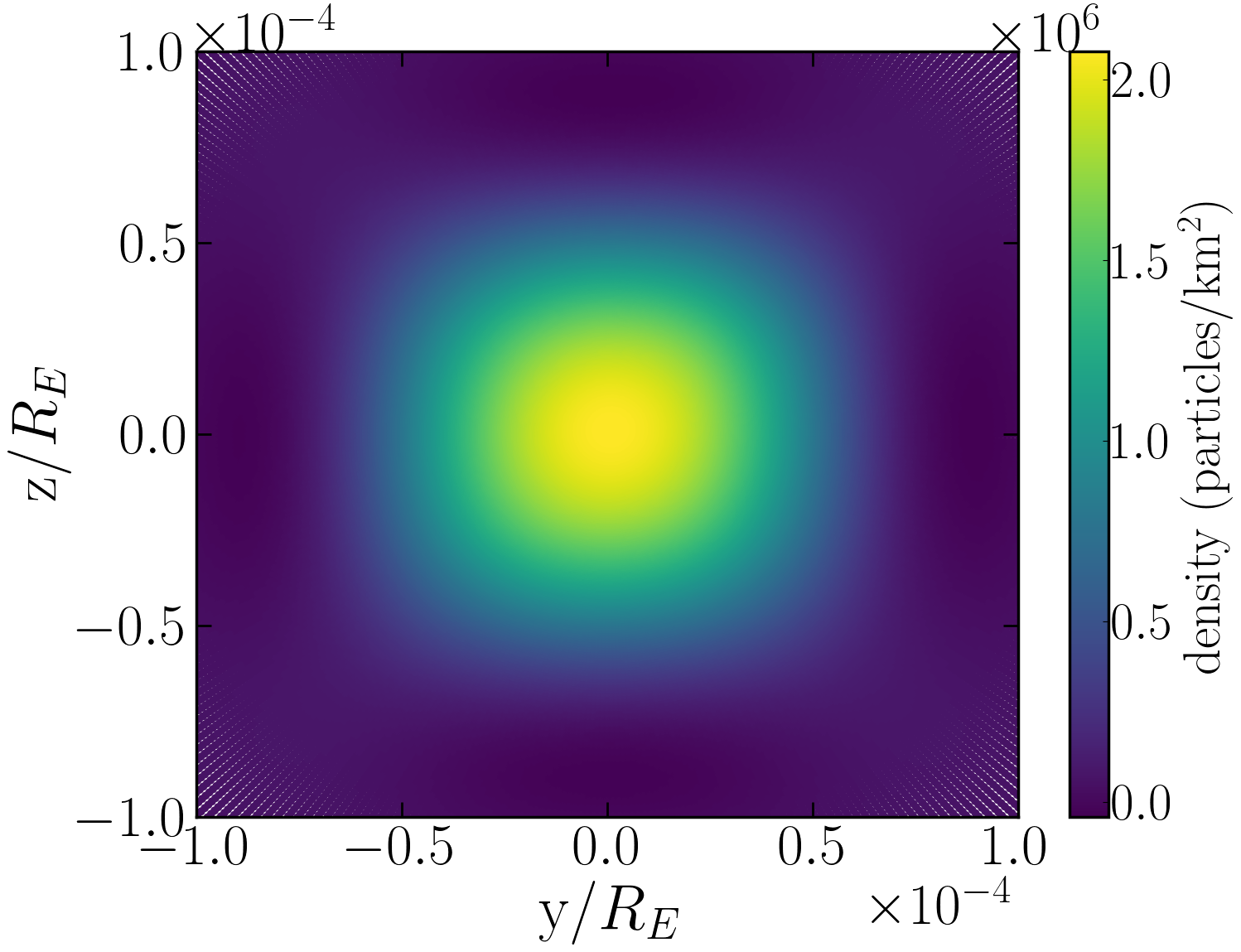}
\caption{
An example of a stream with lab velocity of 20 km/s moving towards the positive X direction and experiencing gravitational self-focusing by the Earth. The top plot shows the XY projection profile of several particles from the same stream focusing at some distance away from the Earth. The YZ profile is evaluated at the focus region. The construction of this particular YZ profile involes 160000 DM particles with impact parameters between 0 and $0.1\,R_E$. Spherical aberration is observed in both top plot where convergence is not at one point, and also bottom plot where density map is not a singular point. The larger the impact parameter of a particle, the bigger the aberration, which has been noticed also in Refs~\cite{Prezeau:2015, sofue:2020}. } 
\label{fig:focusing}
\end{figure}

In this section we calculate the DM density enhancements in the context of small stream dispersion. 
Initially, we discuss the density amplification for single streams. 
The incoming direction of the stream is defined as the positive X-axis, with Y and Z being perpendicular to it. The origin of the reference frame is at the center of the Earth. Figure~\ref{fig:focusing} (top and bottom) shows the simulated XZ and YZ profiles, respectively, illustrating the focusing effect of the gravitational influence of the Earth on DM particles as they traverse through its interior~\cite{prem:1981}. This gravitational focusing leads to the formation of a caustic-like DE region at the focal region.  In an ideal scenario, the particles would converge to a single point, but spherical aberrations as observed in Fig.~\ref{fig:focusing} (bottom) (also discussed in Refs.\cite{Prezeau:2015, sofue:2020}) limit this.


\begin{figure}[h!]
\centering
\includegraphics[width=1\linewidth]{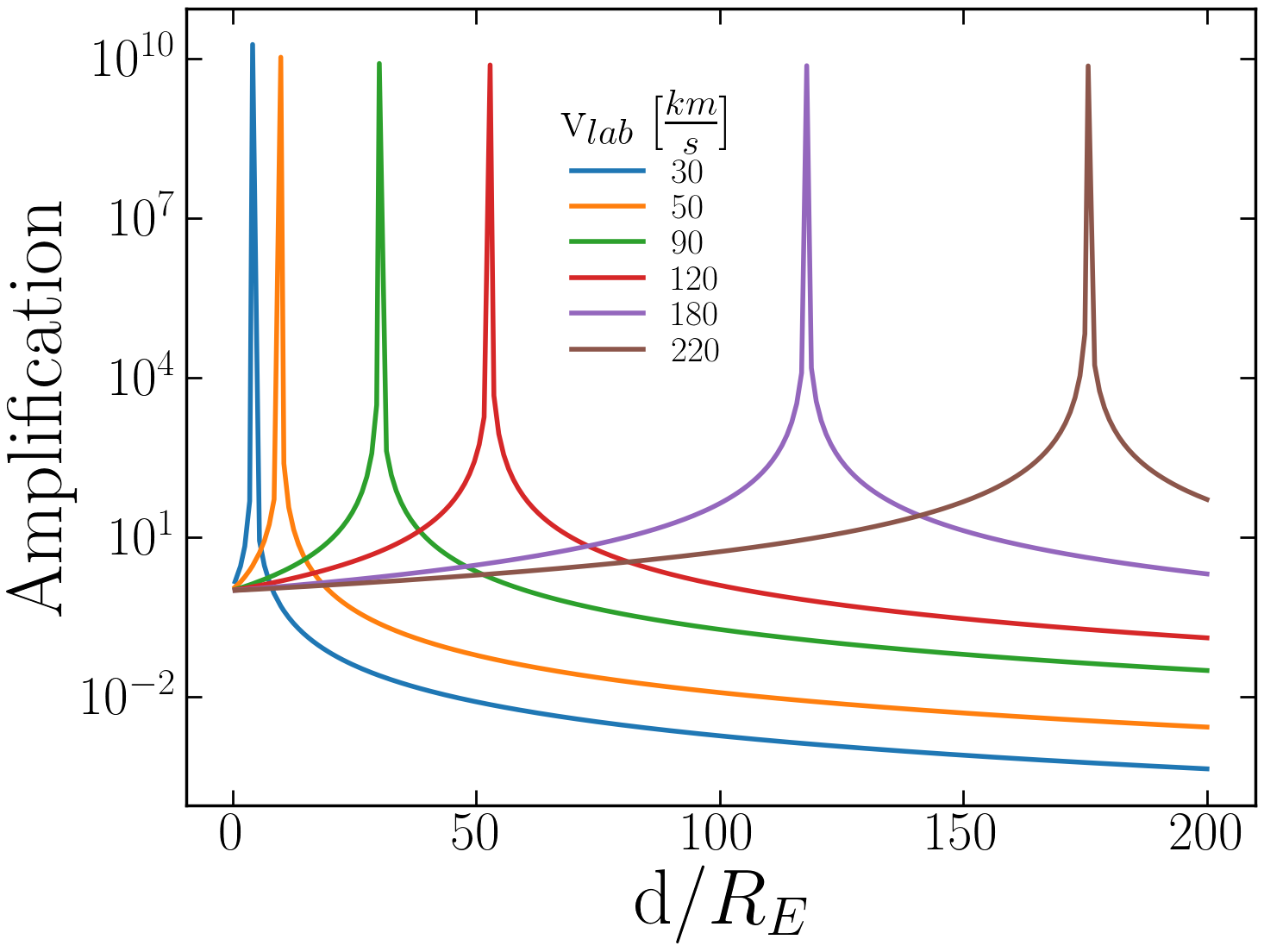}
\caption{DM flux amplification as a function of the radial distance to the center of the Earth in Earth radii for different velocities. The streams in this case are all assumed dispersionless. The amplification peaks at about $10^9$ with a peak height nearly independent on velocity. However, the location of the focus point is strongly dependent on velocity. }
\label{fig:ampl_vs_d}
\end{figure}

Assuming that the velocity of DM particles does not change appreciably while traversing through the Earth, the flux amplification factor is approximately given by~\cite{sofue:2020}: 
\beq
\text{Amplification} \approx \frac{S_0}{S_f} =\frac{\pi r_0^2}{\pi r_{\text f}^2},
\eeq  
where $S_0 =\pi r_0^2$ represents the collection area of DM particles with an impact parameter $r_0$ at the incidence location. Similarly, $S_f =\pi r_{\text f}^2$ denotes the collection area when following the same DM particles to the focus point on the opposite side of the Earth.
In Fig.~\ref{fig:ampl_vs_d}, the DM flux amplification factor is plotted as a function of the radial distance from the Earth's center, with its value being dependent on the velocity of the stream. The depicted six streams are assumed to be dispersionless, and the peak amplification occurs at about $10^9$ (consistent with findings in Refs.~\cite{Prezeau:2015, sofue:2020}).

\begin{figure}[h!]
\centering
  \includegraphics[width=1\linewidth]{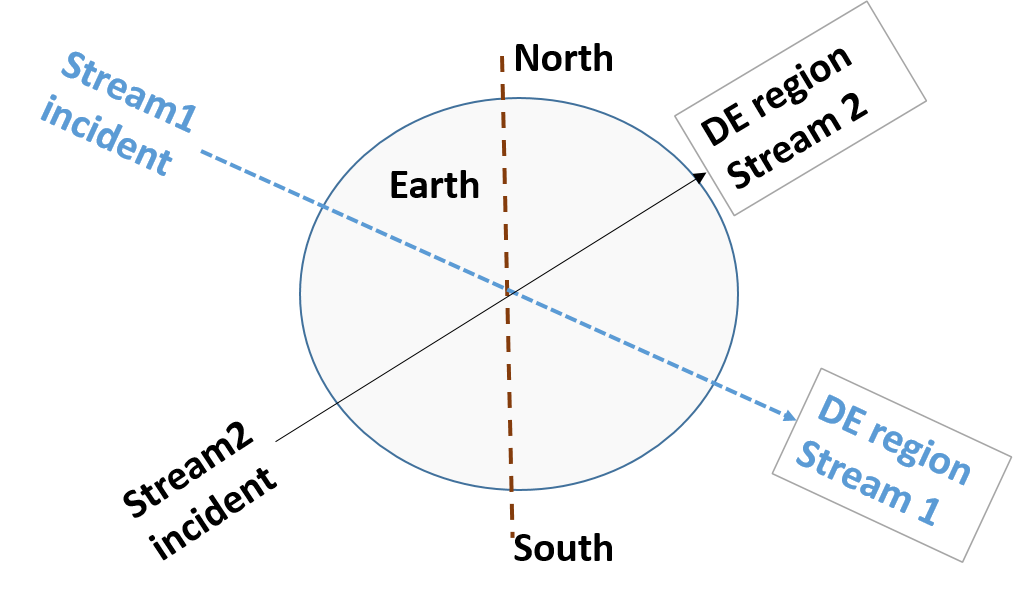}
\caption{Schematics for two DE regions of two streams at different incident velocity and direction. The schematics illustrates how the arrival direction of particles from one stream forms the DE region on the opposite side of the Earth.}
\label{fig:causticregions}
\end{figure}

The density of DM within the DE region (at the focal region) is spatially non-uniform. The highest density is observed at the focal region, gradually decreasing as one moves away from it, as depicted in Figs.~\ref{fig:focusing} and~\ref{fig:ampl_vs_d}.
Given this non-uniformity, the choice of the size and shape of the DE region around the focal region is somewhat arbitrary. We consider several cylindrical sizes with amplification averages ranging from 10 to $10^8$, wherein lower amplification average corresponds to larger size DE region and vice versa. Table~\ref{tab:dimensions_table} provides the dimensions of the DE regions for the various amplification averages. For instance, defining a cylindrical shaped DE region with a length of about $1\,R_E$ and an average radius of about 13 km yields an average amplification of $10^4$. Similarly, a DE region characterized by a cylindrical shape with a length of $0.2\,R_E$ and an average radius of 2.5 km has an average amplification of $10^8$. The radial location of the cylinders depends on the velocity of the stream (see Fig.~\ref{fig:ampl_vs_d}). Fig.~\ref{fig:causticregions} illustrates visually these DE cylindrical shape regions.

\begin{table}
	\centering
	\caption{Derived parameters for DE regions for different streaming DM flux amplifications.}
	\label{tab:dimensions_table}
	\begin{tabular}{l||cccccccc|} 
		\hline
        Amplification & 10 & $10^2$ & $10^3$ & $10^4$ & $10^6$ & $10^7$ & $10^8$\\
        Average &  &  &  &  &  &  &  \\ 
		\hline
		Length (L) [$R_{E}$] & 20.3 & 9.3 & 3.0 & 1.0 & 0.76 & 0.4 & 0.2 \\ 
        \hline
		Width Radius ($r_{\text{ave}}$) [km] & 47 & 46 & 32 & 13 & 9.5 & 5 & 2.5 \\ 
		\hline
	\end{tabular}
\end{table}

\vspace{3mm}
{\bf {Velocity Dispersion Effect}}
\vspace{1mm}

\begin{figure}[h!]
\includegraphics[width=1\linewidth]{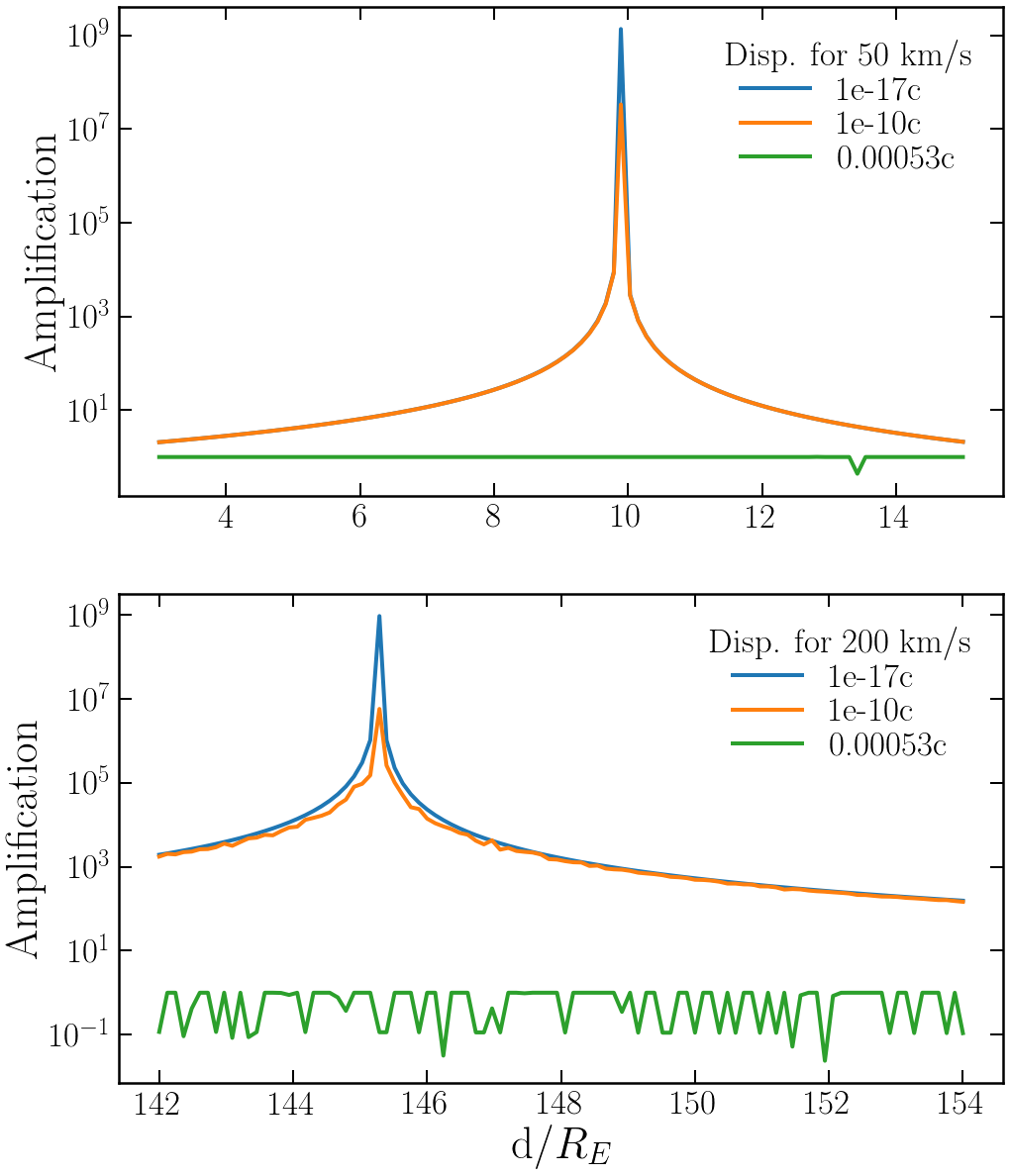}
\caption{The amplification for incident DM velocities of 50 km/s (top) and 200 km/s (bottom) as a function of the radial distance from the center of the Earth. The chosen three values for the dispersion correspond to primordial axions, WIMPs, and the smooth SHM. For the Maxwellian (green) case, the large velocity dispersion is behind the observed amplification values below 1 and for the fluctuations. }
\label{fig:ampl_vs_disp}
\end{figure}

The analysis in Fig~\ref{fig:ampl_vs_d} pertains to dispersionless streams. However, the actual dispersion included in our simulation. Figure~\ref{fig:ampl_vs_disp} describes the amplification as a function of the distance to the center of the Earth for two stream velocities centered at 50 km/s and 200 km/s. Such particles are injected with different velocity dispersion values. Here, we use $\delta v=10^{-17}$c, and $\delta v=10^{-10}$c corresponding to primordial fine grained axion streams and WIMPs, respectively. As a cross-check, we simulated also the case of $\delta v \sim 0.00053$~c for the smooth DM SHM. The dispersion is negligible for primordial axion streams with the amplification factor reaching up to $10^9$, just as in the dispersionless case. This is of potential importance for DM axion detection.  In case of WIMP streams the amplification factor is about a factor of 10 lower than for the fine grained axion streams due to their slightly larger velocity dispersion. For the smooth SHM, as expected, the GF effects become negligible. The noise-like structure in the case of the smooth SHM is attributed to the large dispersion where the amplification is very sensitive in the random selection of input parameters in the simulation.   




\vspace{3mm}
{\bf {Multi-stream density enhancement}}
\vspace{1mm}

Realistically, more than one stream overlaps at a particular location, and the local nominal DM density is the superposition of all contributing streams. Assuming that all DM comes from the sum of $\sim10^{12}$ streams, the density of DM before GF is given as
\beq
\rho_0 = \sum_{k} \rho_{{sk}},
\label{eq:density}
\eeq
where $\rho_{{sk}}$ represents the density contribution to the nominal DM density from the $k$-th stream. When taking into account the gravitational self-focusing by the Earth (Earth mainly causes density enhancement as discussed in preceding section) Eq.~\ref{eq:density} reads as
\beq
\rho =\sum_{k} {A}_{k} \rho_{{sk}},
\label{eq:densitygf}
\eeq
where the amplification factor is specified as $A_{k}$. After gravitational self-focusing by the Earth, to sum up the density of the individual streams, the amplification coefficients are obtained from simulation of the overlapping streams. In locations where there is no dominant term in Eq.~\ref{eq:densitygf}, the contribution of many streams would average to $\rho \sim \rho_{0}$. However, if there is a dominant term Eq.~\ref{eq:densitygf} could be written:
\beq
\rho= A_{j} \rho_{{sj}} + \sum_{k\neq j} A_{k} \rho_{{sk}},
\label{eq:density2parts}
\eeq 
where the $j$-th stream has been singled out from the rest. Since the focus location is strongly dependent on stream velocity (Fig.~\ref{fig:ampl_vs_d}) it is reasonable to assume DE locations where the amplification from a particular $j$-th stream dominates over the rest of the streams (i.e., $A_{j}\gg A_{k}$).
Additionally, if the density after amplification from one stream dominates over the others (i.e., $A_{j} \rho_{sj}\gg \rho_0$), then the second term in Eq.~\ref{eq:density2parts} is smaller than the first and Eq.~\ref{eq:density2parts} reads as:
\beq
\rho \sim A_{j} \rho_{sj}
\label{eq:densitygfunique}
\eeq
The local DM density ratio over the nominal value $\rho_0$ is then:
\beq
\rho/\rho_0 \sim A_{j} \rho_{sj}/\rho_0
\label{eq:densityratio}
\eeq
For DE regions, the simple expressions given by Eqs.~\ref{eq:densitygfunique} and~\ref{eq:densityratio} apply. 

 

\section{Results - Discussion}
\label{sec:discussion}

For DM detection, key factors include understanding the expected stream and DE region count in the vicinity of the Earth, quantifying density enhancements compared to local DM density, and determining their duration. 
The stream quantities with a specific density $\rho_{s}$ are taken from Ref.~\cite{Vogelsberger:2011}. Table~\ref{tab:stream_table} gives a summary of the abundance of the fine grained streams at the solar system and the probability that such streams exist in the solar system vicinity. The stream count is calculated following the approximate relation~\cite{Vogelsberger:2011}:
\beq
N_{s}\cdot \frac{\rho_s}{\rho_0} \sim F(>\rho_s) N_{\text{ts}},
\eeq
where $F(>\rho_s)$ is the fraction of streams with density above $\rho_s$. $N_{s}$ is the number of streams for the specific density, and $N_{\text{ts}}$ is the total number of streams~\cite{Vogelsberger:2011}.  For instance, referencing Table~\ref{tab:stream_table}, a stream with a density of $\rho_s=0.01\,\rho_{0}$ has a 20\% likelihood of being present in our vicinity. Meanwhile, a stream with density $\rho_s=10^{-7}\rho_{0}$ has a 100\% probability, resulting in an estimated stream count of $N_s\sim 2\times 10^6$.

\begin{table}
	\centering
	\caption{DM streams with different densities and their probability to be in solar neighbourhood before any gravitational effects by the solar system bodies set in (see Fig. 10 in ~\cite{Vogelsberger:2011}.)}
	\label{tab:stream_table}
	\begin{tabular}{lcr} 
		\hline
		Stream density ($\rho_{0}$) $\vert$ & Stream Count $\vert$ & Probability (\%)\\
		\hline
		1 & 1 & 0.002\\
		0.1 & 1 & 0.2\\
		0.01 & 1 & 20 \\
		10$^{-3}$ & 10 & 100\\
		10$^{-4}$ & 500 & 100\\
		10$^{-5}$ & $2\times10^4$ & 100 \\
        10$^{-6}$ & $4\times10^5$  & 100 \\
        10$^{-7}$ & $2\times10^6$  & 100 \\
		\hline
	\end{tabular}
\end{table}

\begin{figure}[h!]
\includegraphics[width=1\linewidth]{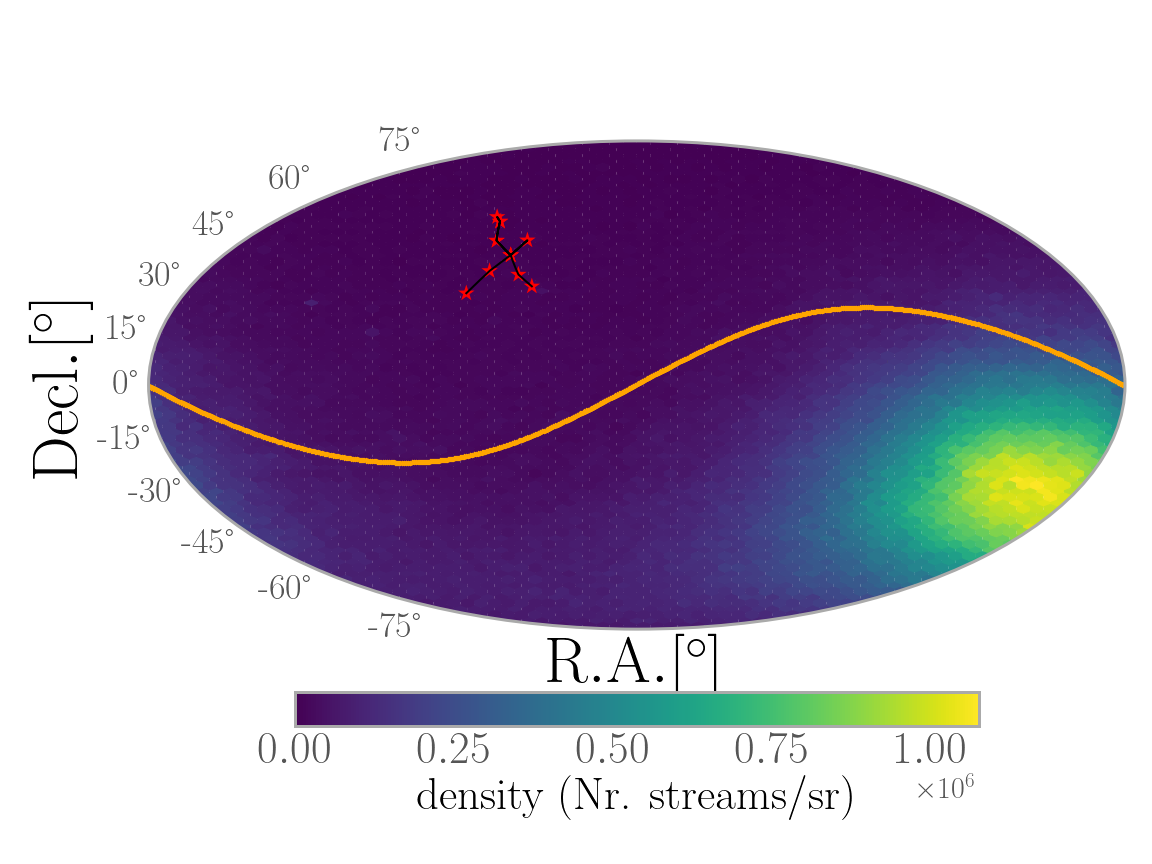}
\caption{
Equatorial coordinates map distribution of DM stream directions of all velocities.  
For orientation reference, the Cygnus constellation is denoted by red stars, and the solar ecliptic for a period of one year is marked by an orange line. There appears to be a significant concentration of stream directions on the opposite side of Cygnus. This is consistent with more streams being expected to arrive from Cygnus direction and exiting on the opposite side of the Earth. This excess is caused by the kinematical boost of the Sun's motion toward Cygnus.  
}
\label{fig:arrivalall_1}
\end{figure}

\begin{figure}[h!]
\includegraphics[width=1\linewidth]{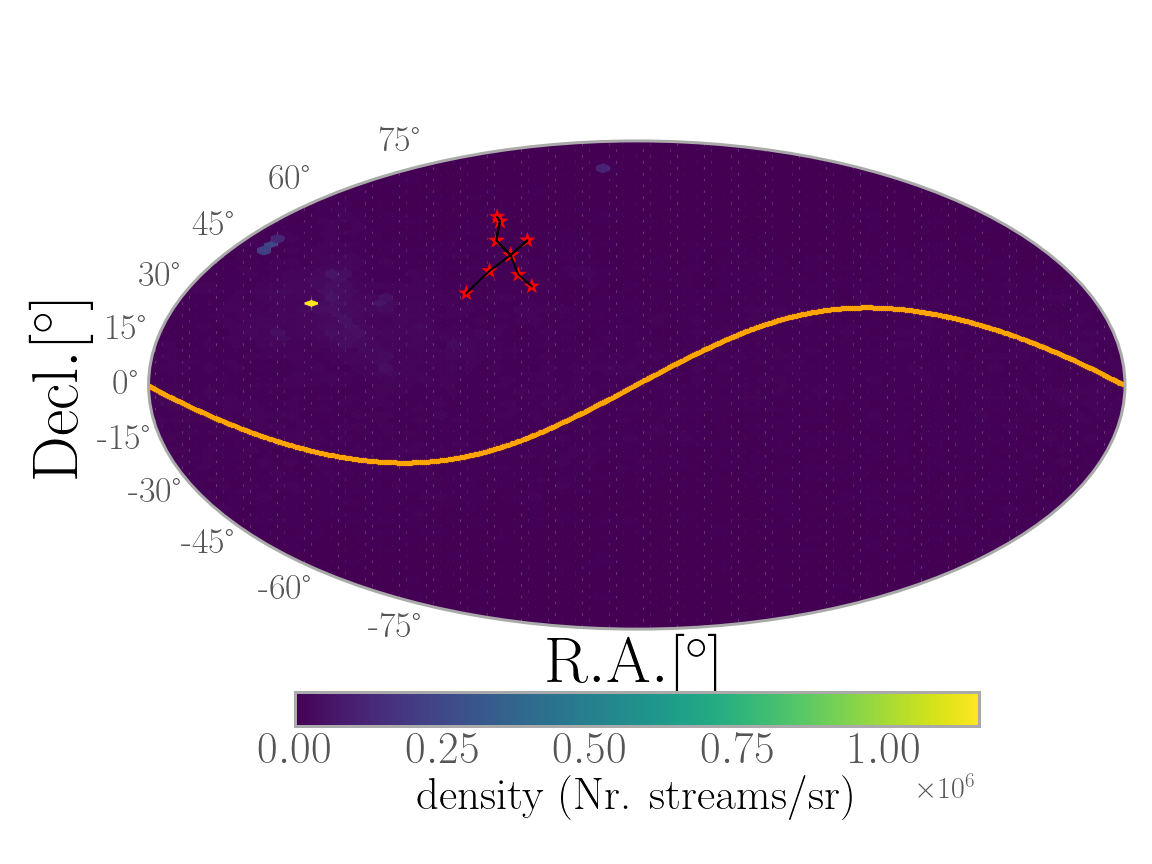}
\caption{
Equatorial coordinates map distribution of stream directions for DM streams with velocity less than 50 km/s. The distribution is approximately isotropic for low velocities.  For orientation reference, the Cygnus constellation is depicted with red stars and the ecliplic with orange circles.}
\label{fig:arrivalall_2}
\end{figure}

In order to examine whether specific geographic locations are at more advantageous position for detection, we analyze the direction of the DM streams as they exit the Earth, and consequently their DE regions as illustrated in Fig.~\ref{fig:causticregions}. 
The angular distribution of stream directions in terms of equatorial coordinates: right ascension and declination is shown in Figs.~\ref{fig:arrivalall_1} and ~\ref{fig:arrivalall_2}. To provide an orientation aid, the orange line delineates the ecliptic plane for a period of one year, while red stars indicate the right ascension and declination of the Cygnus constellation.

Figure~\ref{fig:arrivalall_1} shows the angular distribution of stream directions for all stream velocities. The kinematical boost due to Sun's motion towards the direction of Cygnus leads to an excess of stream directions op the opposite side of Cygnus as expected (see Ref.~\cite{chare:2018}). The majority of the streams represented in Fig.~\ref{fig:arrivalall_1} result in DE regions at considerable distance from the Earth's surface. In contrast, Fig.~\ref{fig:arrivalall_2} provides a similar angular distribution of stream directions but it considers only streams with velocities below 50 km/s, since for these streams the corresponding DE regions are in close proximity to the Earth. At lower velocities, the Sun's gravitational influence significantly alters the stream directions (see Eq.~\ref{eq:velocity}), resulting in a more uniform distribution. Consequently, the anisotropy observed in Fig.~\ref{fig:arrivalall_1} vanishes in Fig.~\ref{fig:arrivalall_2}. This suggests that for low-velocity streams, the distribution of stream directions, and consequently, the DE regions is isotropic.


The results of Fig.~\ref{fig:arrivalall_2} lead us to conclude that DE regions in the vicinity of the Earth are equally likely to be encountered in experiments at any geographical location on Earth. Note, that each data point in these plots represents a stream direction as they exit the Earth, not DM particle densities.

Next, we derive the number of DE regions for Earth bound experiments by choosing two different distances from the center of the Earth $d_{\text{obs}} \sim 1\,R_{E}$ (Earth's surface) and $d_{\text{obs}}\sim 1.5\,R_{E}$ (in orbit)\footnote{The choice of $d_{\text{obs}}\sim 1.5\,R_{E}$ is somewhat arbitrary, different enough from Earth's surface, yet within feasibility of satellite locations. }.
We introduce the concept of a cutoff velocity for a given stream, denoting the upper limit at which its particles can be accommodated within the specified DE region. It is evident that larger DE region sizes result in higher cutoff velocities. The values of these cutoff velocities for the two aforementioned Earth observation radial distances and various amplifications are provided in Table~\ref{tab:vcut_table}.  For a DM detector located at the Earth’s surface, the cutoff DM velocities fall within the range of $75$~km/s and $10$~km/s, for DM flux enhancements between 10 and $10^8$ , respectively. Likewise, as given in  Table~\ref{tab:vcut_table}, for a DM detector positioned at about $1.5\,R_{E}$ from the center of the Earth, the corresponding cutoff velocities are $78$~km/s and $14$~km/s. 



\begin{table}
	\centering
	\caption{The cutoff velocities for DE regions for different DM flux amplifications and different observation distances.}
	\label{tab:vcut_table}
	\begin{tabular}{l||cccccccc|} 
		\hline
		Amplification & 10 & $10^2$ & $10^3$ & $10^4$ & $10^6$ & $10^7$ & $10^8$\\
        Average &  &  &  &  &  &  &  \\ 
		\hline
		$v_{\text{cut}}\leq$ (km/s) & 75 & 51 & 29 & 18 & 16 & 13 & 10 \\ 
        ($d_{\text{obs}} \simeq 1\,R_{E}$ ) &  &  &  &  &  &  &  \\ 
		\hline
		$v_{\text{cut}}\leq$ (km/s)  & 78 & 52 & 32 & 22 & 19 & 17 & 14 \\ 
         ($d_{\text{obs}} \simeq 1.5\,R_{E}$ )  &  &  &  &  &  &  &  \\
		\hline
	\end{tabular}
 
\end{table}

\begin{figure}[h!]
\includegraphics[width=1\linewidth]{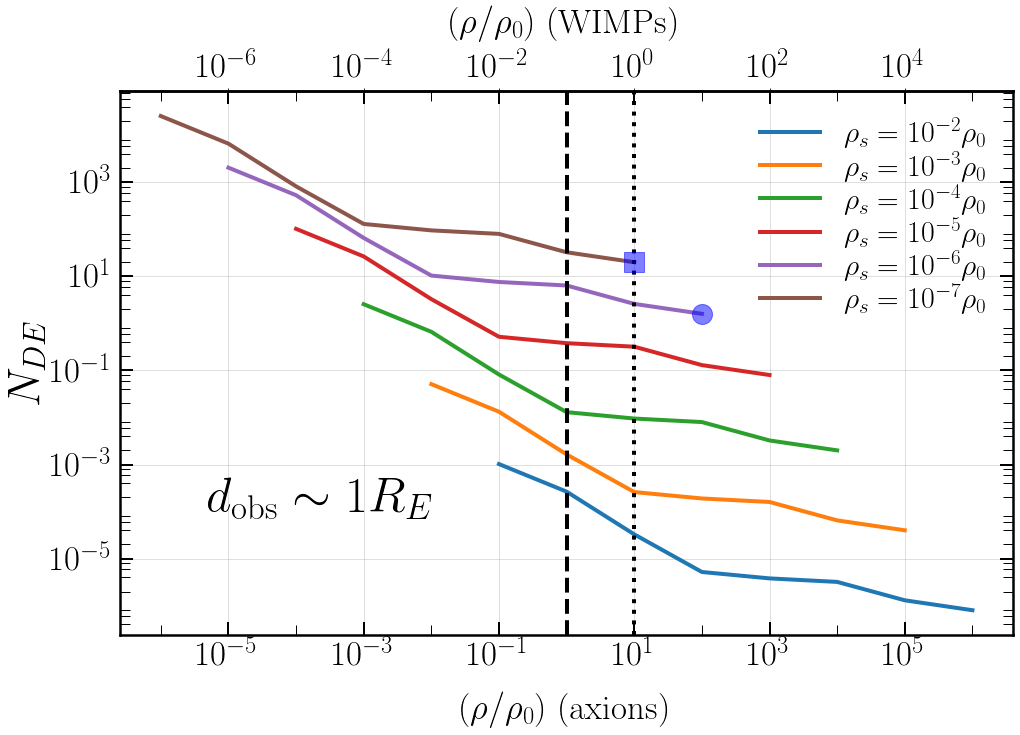}
\caption{
Simulation: The number of DE regions expected at the Earth's surface ($d_{\text{obs}}\sim 1\,R_{E}$). Each line represents different incident stream density as taken from Table~\ref{tab:stream_table}. A line spans a range in the x-axis of 7 orders of magnitude because of the built in DM flux amplification between 10 and $10^8$. The x-axis gives the ratio of the DE density over the nominal DM density $\rho_0$, as described by Eq.~\ref{eq:densityratio}. The density enhancement ratio for axions is given at the bottom x-axis and for WIMPs at the top. 
The parameter space to the right of the dashed vertical line indicates DE densities greater than $\rho_0$ for axions, and similarly to the right of dotted vertical line for WIMPs. The square marker corresponds to an incident stream density of $\rho_s \sim 10^{-7}\rho_0$, where from $5\times 10^6$ axion streams about 20 DE regions with density ratio $\rho/\rho_0 \sim10$ are expected; and the circle marker corresponds to incident axion streams with density of $\rho_s \sim 10^{-6}\rho_0$, where from $4\times 10^5$ axion streams, about 1.6 DE regions with density ratio $\rho/\rho_0 \sim 100$ are expected.
}
\label{fig:stream_sensitivity1}
\end{figure}

\begin{figure}[h!]
  \includegraphics[width=1\linewidth]{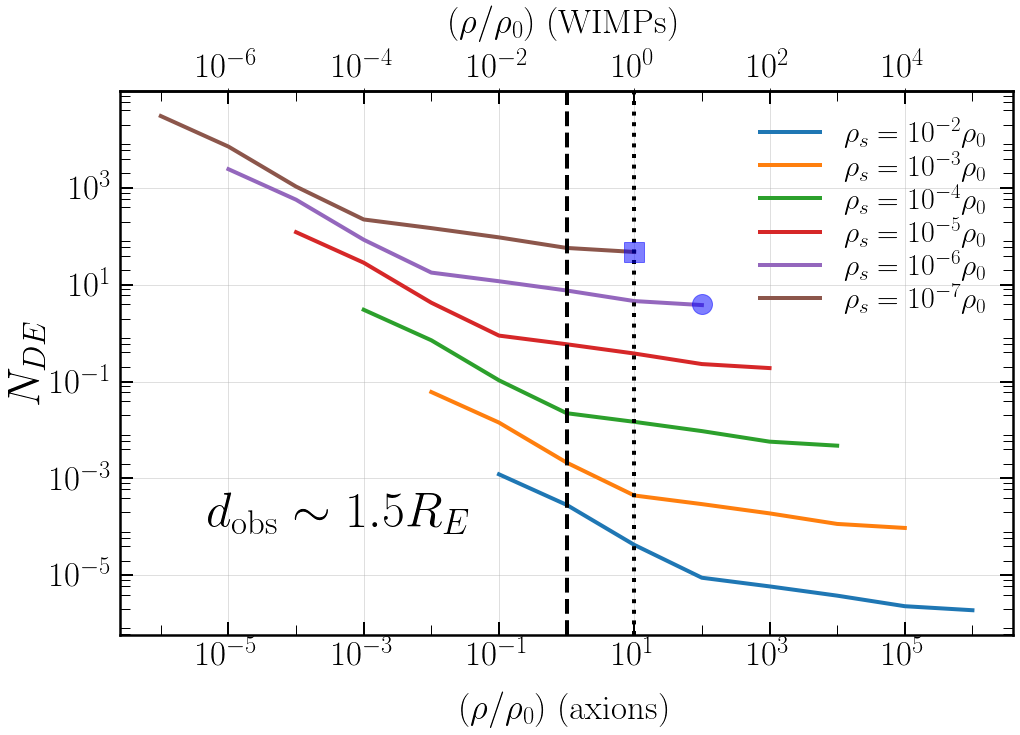}
\caption{
Simulation: The number of DE regions in near Earth surface ($d_{\text{obs}}\sim 1.5\,R_{E}$). Each line represents different incident stream density as taken from Table~\ref{tab:stream_table}. A line spans a range in the x-axis of 7 orders of magnitude because DM flux amplification is between 10 and $10^8$. The x-axis represents the ratio of the DE density over the nominal mean DM density $\rho_0$ as described by Eq.~\ref{eq:densityratio}. The amplification factor for the axions is given in the bottom x-axis and for WIMPs at the top. The parameter space to the right of the dashed vertical line indicates DE densities greater then $\rho_0$ for axions, and similarly to the right of dotted vertical line for WIMPs. The square marker corresponds to an incident stream density of $\rho_s = 10^{-7}\rho_0$, where from $5\times 10^6$ axion streams, about 48 DE regions with density ratio $\rho/\rho_0 \sim 10$ are expected; and the circle marker corresponds to incident axion streams with density of $\rho_s = 10^{-6}\rho_0$, where from $4\times 10^5$ axion streams, about 4 DE regions with density ratio $\rho/\rho_0 \sim 100$ are expected.
}
\label{fig:stream_sensitivity5}
\end{figure}

Figure~\ref{fig:stream_sensitivity1} and~\ref{fig:stream_sensitivity5} give the number of DE regions at $\sim 1\,R_{E}$ and $\sim\,1.5 R_{E}$ as a function of density ratio $\rho/\rho_0$. Each line on the plot is generated assuming a constant stream density chosen from Table~\ref{tab:stream_table}. According to Eq.~\ref{eq:densityratio}, even for constant incident stream density, the final density ratio of the DE regions is dependent on the DM flux amplification values ranging from 10 to $10^8$ which determine also the lower and upper limits for each line on the plots. The x-axis denotes the density ratio for axions at the bottom and for WIMPs at the top.
\footnote {We use the expression from equation~\ref{eq:densitygfunique} to calculate the density ratio. We emphasise that this expression works in a multi-stream environment as long as $A_{j}\gg A_{k}$, and $A_{j} \rho_{sj}\gg \rho_0$ which would be fine for density ratios greater than $\sim 10$; the graphs of Figs.~\ref{fig:stream_sensitivity1}, ~\ref{fig:stream_sensitivity5} are then a better approximation of the multi-stream configuration. } 




For instance, in Fig.~\ref{fig:stream_sensitivity1}, the square marker corresponds to incident streams with $\rho_s=10^{-7}\rho_0$. In this scenario, given a total of $5\times 10^6$ axion streams, 20 DE regions would be expected at the Earth's surface, each with the density ratio $\rho/\rho_0 \sim 10$. On the other hand, the circular marker pertains to incident streams with $\rho_s=10^{-6}\rho_0$. In this configuration, given $4\times 10^5$ axion streams, an average of 1.6 DE regions are expected at the Earth's surface, each with the density ratio $\rho/\rho_0 \sim 100$. For the observation distance of $\sim 1.5\,R_{E}$ shown in Fig.~\ref{fig:stream_sensitivity5}, the corresponding markers suggest that there would be about 48 axion DE regions with density ratio $\rho/\rho_0 \sim 10$, and about 4 axion DE regions with density ratio $\rho/\rho_0 \sim 100$.

\begin{figure}[h!]
\includegraphics[width=1\linewidth]{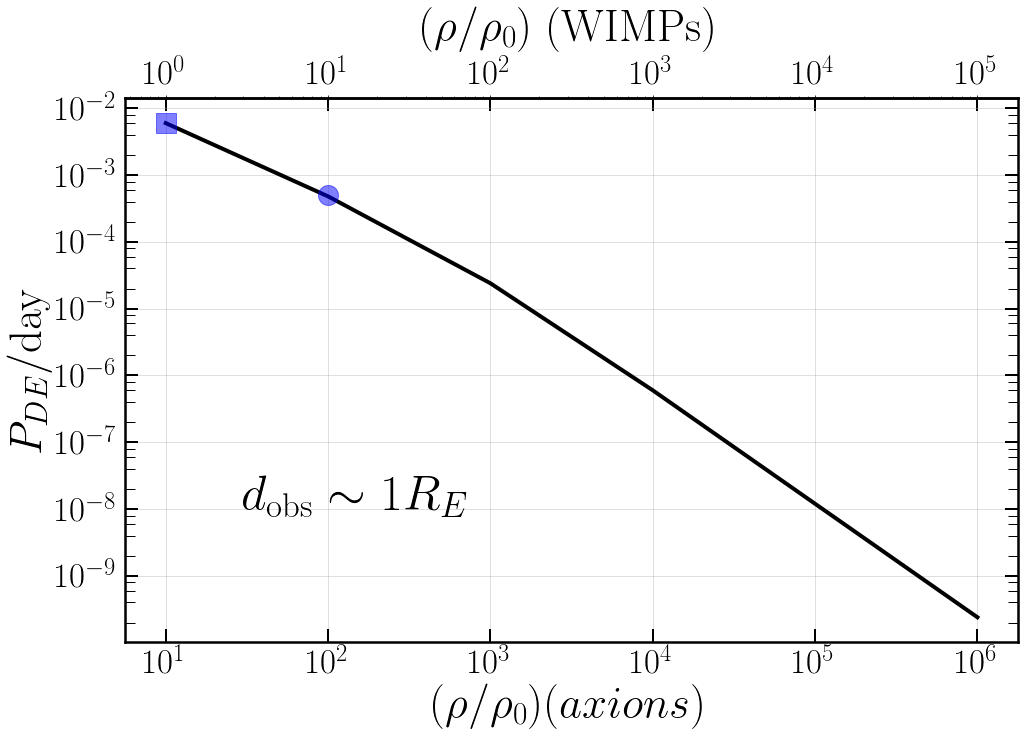} 
\caption{
The probability of encountering a DE region per day for an experiment located at the Earth's surface ($\sim\,1 R_{E}$). The square marker shows that the probability to encounter an axion DE region with density ratio $\rho/\rho_0 \sim 10$ is $\sim 6\times 10^{-3}$,  while the circle marker shows that the probability to encounter an axion DE region with a density ratio $\rho/\rho_0 \sim 100$ is $\sim 5\times 10^{-4}$. }
\label{fig:timesensitivity1}
\end{figure}

\begin{figure}[h!]
\includegraphics[width=1\linewidth]{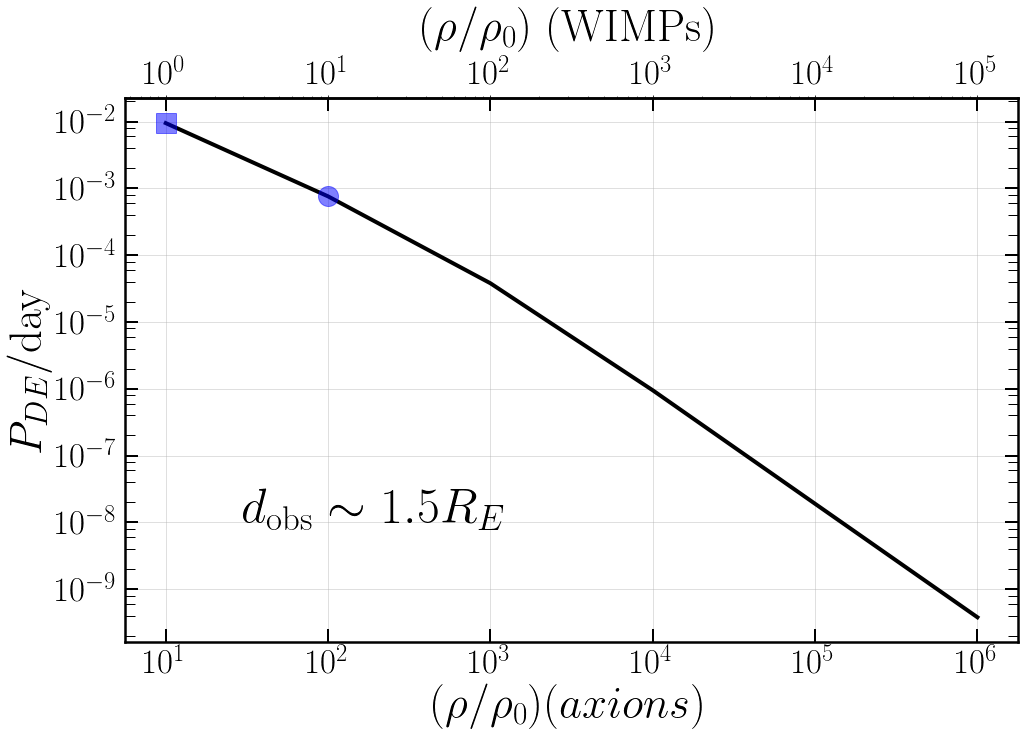} 
\caption{
The probability of encountering a DE region per day for an experiment located at an observation distance of about $\sim\,1.5 R_{E}$. The square marker shows that the probability to encounter an axion DE region with density ratio $\rho/\rho_0 \sim 10$ is $\sim 1\times 10^{-2}$,  while the circle marker shows that the probability to encounter an axion DE region with a density ratio of $100$ is $\sim 8\times 10^{-4}$.}
\label{fig:timesensitivity5}
\end{figure}

Figures~\ref{fig:stream_sensitivity1} and ~\ref{fig:stream_sensitivity5}  show possible DE regions with different stream densities assuming full sky coverage. The experiment's field of view (f.o.v.)  moves due to Earth's motion. Experiments require an understanding of the probability of detecting a DE region per day, accounting for the specific density and duration of the anticipated transient signature. The probability of encountering one appropriate DE region per day is given by:
\beq
\label{eq:timeEqn}
P_{DE}/\text{day} \sim N_{DE} N_{\text{det}} P_{\text{loc}},
\eeq
where $N_{{DE}}$ is the number of DE regions present. For the observation distances mentioned above, they are obtained from Figs.~\ref{fig:stream_sensitivity1} and ~\ref{fig:stream_sensitivity5}. $N_{\text{det}}$ is the number of detectors, and $P_{\text{loc}}$ is the solid angle distribution probability of DE regions in space. For small velocities, it is shown in Fig.~\ref{fig:arrivalall_2} that the expected probability distribution of DE regions is approximately isotropic. The distribution probability is then calculated as
\footnote{Size of the detector is assumed to be much smaller than size of DE region.}
\beq
\label{eq:probability}
P_{\text{loc}} \sim \max \left (\frac{2\pi d_{\text{obs}} \text{cos} (\phi) (2 r_{\text{ave}})}{4 \pi d^{2}_{\text{obs}}}, \frac{\Delta\Omega}{4\pi}\right),
\eeq
where $r_{\text{ave}}$ is the average radius of the cylindrically-shaped DE region for a particular DM flux amplification factor given by Table~\ref{tab:dimensions_table}, $\phi$ is the latitude of the detector, and $\Delta\Omega$ is the solid angle f.o.v. of the experiment during one day. The numerator represents the DE region/f.o.v. area covered during one day, and the denominator gives the $4\pi$ coverage. Since the f.o.v. is experiment dependent, in what follows, the left term in Eq.~\ref{eq:probability} is used as baseline to calculate the probability. Figures~\ref{fig:timesensitivity1} and ~\ref{fig:timesensitivity5} give the probability to encounter one DE region per day, for one single detector module for a mid latitude observer located at about $\phi\sim 40^{o}$.  

As an example, the probability to encounter one axion DE region per day, for one experiment at the Earth's surface ($1\,R_{E}$), with a density ratio $\rho/\rho_0 \sim 10$ is $\sim$0.006. The same probability for an axion density ratio of $100$ is $\sim 5\times 10^{-4}$. The corresponding values for one experiment at $\sim 1.5\,R_{E}$, and an axion density ratio $\rho/\rho_0 \sim 100$ becomes $8\times 10^{-4}$. 

The densest incident streams could yield DE regions with density ratio of up to $10^6$ after GF; however, the probability to encounter one such a region per day is suppressed due to the scarcity of these streams (the probability is below $10^{-9}$ for both observation distances mentioned here). Even though, the probability is small for these rare events, it is worth noting the steadily increasing number of DM experiments worldwide.  


The detector moves with respect to the DE regions as the Earth's rotates giving rise to transient signals. Of note, a detector at the surface of the Earth moves at a speed of $\sim 0.5\, ({\text {km/s}}) \, \text{cos}(\phi)$. The transient signal duration for an experiment that is moving through a DE region at an observation distance $d_{\text{obs}}$, is given by
\beq
t_{\text{enc}} \sim \frac{2\, r_{\text{ave}}}{ (d_{\text{obs}}/R_{E})\, 0.5\, \text{cos} (\phi)}
\eeq  
For $r_{\text{ave}}=2.5\, \text {km}$, the transient time is $\sim$13 seconds at the Earth's surface ($1\, R_E$), and $\sim$9 seconds at ($1.5\,R_{E}$).

We stress that probabilities derived by Eq.~\ref{eq:timeEqn} and shown in Figs.~\ref{fig:timesensitivity1} and~\ref{fig:timesensitivity5} express the lack of knowledge on the position of a DE region. However, if an experiment happens to be on the propagating path of a DE region, the density transient of about 10 seconds per day will appear consistently at the same sidereal time during the period of days to weeks or even longer until the DE region moves out of the f.o.v. of the experiment.\footnote{ The uncertainty in the duration of persistence from days to weeks is dependent on the f.o.v. and sensitivity of a particular detector/antenna.} This is an important new input for streaming DM signal identification that arises from our simulation. 

Evidently, a network of  many experiments appears promising, particularly as long as the mass of the DM particles and their interaction strength or their stream configuration are not known. 
For instance the GNOME network~\cite{gnome:2021} is designed to search for transient DM signals using magnetometers spread out across the world, the electric dipole moments (EDM) storage ring experiments~\cite{yannis:2018,yannis:2021} with couple sites, and the ECHO idea~\cite{ariel:2019}, etc., fit the derived results. The ECHO method explores the use of Earth bound radio telescopes for axion searches, and has provided encouraging sensitivity results for streaming DM. The results are described in Ref.~\cite{echostreams:2022}. 

Moreover, experiments at about the same latitudes can correlate time-delayed signals, further improving the signal identification. Note that experiments at the same latitudes would encounter the same DE region at different times because of tangential movement of an Earth bound experiment as the Earth rotates.

Another possible implication of the GF effect of streaming DM constituents by the Earth, results in possible overlap with conventional cosmic ray studies. Because, the appearance of caustics-like shapes downstream of an incident low-speed DM stream might mimic upward escaping cosmic-ray like events from the Earth's surface. In this context we mention the intriguing anomalous events observed by the Antarctic Impulse Transient Antenna (ANITA)~\cite{Anita:2016,Anita:2018,Anita:2020} that appear as energetic cosmic showers emerging from the Earth's surface.  Interestingly, the advocated anti-quark nugget (AQN) model by Zhitnitsky~\cite{Ariel_Zh:2003}, which has the potential to solve the DM problem, was proposed as an explanation for these apparently anomalous events~\cite{Ariel_Zh:2022}. 

Furthermore, we wish to stress that the simulation tools developed in this work can accommodate any DM particle. \footnote{The simulation code can be provided to interested researchers upon reasonable time request. }
Finally, should a DM particle be identified, one can use the same study presented here to eventually derive the properties of the DM distribution in our Galaxy. Our results show that transient DE regions of high density around the Earth would form only if the streaming  nature of DM is true. 

\subsection{Numerical Examples}
In this section, we mention some practical figures of merit that can be used to initiate a search for such DE regions.  

At the surface of the Earth, there is an average of about 1.6 axion DE regions with radius of $\sim 2.5$~km and a length of $0.2\,R_E$, and with a density ratio $\rho/\rho_0 \sim 100$. An experiment at the surface of the Earth has a probability of about $5\times 10^{-4}$ per day of running into such a region as the Earth rotates and the experiment with it. The transit time is determined from the size of the DE region and the Earth's rotation velocity and is about 13 seconds in this case. 
At a distance of $\sim 1.5\,R_{E}$ there are about 4 axion DE regions with radius of $\sim 2.5$~km, length of $0.2\,R_E$, and with a density ratio $\rho/\rho_0 \sim 100$. An experiment at this location has a probability of about $8\times 10^{-4}$ per day to encounter a DE region with a transit time of about 9 seconds. \footnote{The transit time for this case also depends on the relative speed of the space experiment with respect to the Earth.}

\section{Summary - Conclusion}
\label{sec:summary}

This simulation work addresses the gravitational effects by the Sun and the Earth for streaming DM. More specifically, we concentrate on the fine grained streams from Ref.~\cite{Vogelsberger:2011}. The simulations for a streaming DM configuration show that the Sun primarily alters the direction of the Earth bound DM particles from streams, but it does not result in significant increase of their flux at the Earth. Instead, notable flux enhancements can be attributed to Earth's gravitational self-focusing. The peak DM flux amplification at the Earth's surface occurs when DM particles have incident velocities in the range of 10 to 15 km/s.

The primary benefit, if the dark sector is comprised of streams is that, as the result of the Earth's gravitational self-focusing, given the considerable number of these streams, certain streams are likely to create regions of enhanced spatio-temporal density at the Earth's vicinity. Temporal density enhancements of couple orders of magnitude over the nominal DM density $\rho_0$ are feasible. The transient signals have a duration of about 10 seconds per day and persist from days to weeks at a particular location until they move out of the f.o.v. of the detector. Naturally, a network of experiments increases the probabilities of such encounters. Furthermore, experiments at about the same latitudes could result in correlations between signal candidates. 
The expected transients can become a unique and novel signature to unravel the existence of streams and their structure, which in turn would give feedback about cosmological models of DM. Similar transients in exo-solar systems can be explored~\cite{perrymankonstantin:2021}. This indicates a far-reaching perspective, once data of long time series from nearby exo-planetary systems become available.




\appendix
\section{Trajectories in the Field of Sun and Earth}
For verification, DM particle trajectories are simulated in the combined gravitational field by the Sun and the Earth by solving Eq.~\ref{eq:sunearth}. The density enhancement obtained using this approach is not significantly different from the results obtained using Eqs.~\ref{eq:velocity},~\ref{eq:vlab}, and~\ref{eq:Earth}. However, simulations with Eq.~\ref{eq:sunearth} are computationally intensive as the majority of particles fall outside the Earth, given its relatively small aperture compared to the size of the solar system. 
The Newton's Second Law equation guiding DM particles trajectories in the gravitational field by the Sun and the Earth is given by the following:
\begin{widetext}
\beq
\label{eq:sunearth}
{\bf \frac{d{\bf v_{i}}}  {dt}=
\left\{
\begin{array}{ll}
 -\frac{\text{G} \text{M}_{\odot}}{(\text{r}_i - \text{r}_\text{S})^3}(\bf r_i -r_\text{S}) -\frac{\text{G} \text{M}_{{E}}}{(\text{r}_i - \text{r}_\text{E})^3}(\bf r_i -r_\text{E}) & \mbox{outside Sun and Earth},\\[0.5em]
 -\frac{4\pi }{3}\text{G} \eta_\text{S} (\bf r_i -r_S) -\frac{\text{G} \text{M}_{{E}}}{(\text{r}_i - \text{r}_\text{E})^3}(\bf r_i -r_\text{E}) & \mbox{inside Sun, outside Earth},\\[0.5em]
 -\frac{4\pi }{3}\text{G} \eta_\text{E} (\bf r_i -r_E) -\frac{\text{G} \text{M}_{\odot}}{(\text{r}_i - \text{r}_S)^3}(\bf r_i -r_S) & \mbox{inside Earth, outside Sun},\\
\end{array}
\right\}}
\eeq
\end{widetext}
where $\text{r}_\text{E}$, $\eta_\text{E}$, and $\text{r}_\text{S}$, $\eta_\text{S}$, are positions and densities for the Earth and the Sun, respectively.

\acknowledgments

We sincerely thank Mark Vogelsberger for the numerous feedback about setting up simulation inputs for the fine grained streams. Yannis Semertzidis is acknowledged for encouraging discussions about DM searches. AK acknowledges Messiah University Scholarship Programs for support of the work. KZ thanks Antonis Gardikiotis for his help.






\bibliographystyle{JHEP}
\bibliography{main}

\end{document}